# Molybdenum Disulfide Bilayers Hybridized with Reduced Graphene Oxide Nanosheets for Enhanced Hydrogen Evolution Reaction

Bruno Ferreira Brischi,[†,1] Gabriela Frajtag,[†,1] João Pedro da Silva Mariano,[†,1] Laís Fernanda Medeiros Ruela,[†,1] Gustavo Duarte Verçosa,[1] Camila Desiderio Fernandes,[1,2] Murilo Santhiago,[3,4] Letícia Mariê Minatogau Ferro,[5] Leandro Merces[1,*]

[†]*These authors contributed equally to this work.*
[1]*Ilum School of Science, Brazilian Center for Research in Energy and Materials, 13087-548 Campinas-SP, Brazil.*
[2]*Institute of Physics "Gleb Wataghin", University of Campinas, 13083-859 Campinas-SP, Brazil.*
[3]*Brazilian National Nanotechnology Laboratory, Brazilian Center for Research in Energy and Materials, 13083-100 Campinas-SP, Brazil.*
[4]*Postgraduate Program in Nanosciences and Advanced Materials. Federal University of ABC, 09210-580 Santo André-SP, Brazil.*
[5]*Institute of Chemistry, University of Campinas, 13083-970 Campinas-SP, Brazil.*
[*]*E-mail:* leandro.merces@ilum.cnpem.br

**ABSTRACT**

Two-dimensional (2D) molybdenum disulfide ($MoS_2$) nanosheets have attracted attention as a promising and cost-effective alternative catalyst for the hydrogen evolution reaction (HER). However, their aggregation and poor conductivity limit their catalytic activity. Consequently, researchers are exploring ways to improve the conductivity and density of electroactive edges of $MoS_2$ through hybridization with akin advanced 2D materials. Here, reduced graphene oxide (rGO) nanosheets are added to a liquid-phase-exfoliated $MoS_2$ dispersion to create drop-casting hybrid $MoS_2$@rGO electrodes for HER. The findings reveal the formation of ~1.1 nm-thick $MoS_2$ bilayers, with the catalytic performance of $MoS_2$@rGO dependent on the degree of graphene oxide reduction. The rGO moderate reduction prevents the $MoS_2$ bilayers from restacking, improves conductivity, retains oxygenated groups that enhance interlayer spacing, and exposes more electroactive edges. When hybridized with rGO crafted at optimal reduction time, the hybrid system eclipses the efficiency of pristine $MoS_2$ bilayers by attaining the lowest overpotential (~70 mV) and the lowest Tafel slope (~46 mV dec$^{-1}$). This shows that $MoS_2$ bilayers hybridized with rGO offer a promising method to outperform the electrocatalytic efficiency of $MoS_2$-based electrodes. These findings expand opportunities for future strain engineering, defect engineering, and high-end twist-angle assemblies for hybrid systems combining $MoS_2$ bilayers and rGO.

**INTRODUCTION**

Two-dimensional (2D) materials have revolutionized condensed matter research by offering unprecedented opportunities to tailor electronic, optical, chemical, and mechanical properties at the atomic scale. Their atomically thin nature enables stacking without lattice mismatch, allowing the creation of bilayer and multilayer structures with tunable interfacial coupling,[1,2] twist angles,[3,4] and moiré superlattices.[5] These structural degrees of freedom give rise to exotic phenomena,[6–9] expanding the functional landscape beyond conventional bulk materials. Strain engineering further amplifies these effects, enabling control over polarization,[2,10] domain structures,[11,12] optoelectronic responses,[13] and electron-transfer rates[14,15] through nanoscale gradients. Such versatility is critical for the miniaturization of intelligent devices,[16] where integrating sensing,[17] computing,[18] energy harvest[19] and storage[20] within confined spaces is essential. Consequently, the precise control over growth, stacking, interfacial properties and defect engineering of 2D materials remains a central challenge and a gateway to next-generation technologies in electronics,[21] energy,[22] and sustainability.[23]

Molybdenum disulfide ($MoS_2$) is a transition metal dichalcogenide that has garnered attention as a promising catalyst for sustainable hydrogen production due to its abundance, low toxicity, tunable electronic behavior, and facile synthesis.[24–27] The bulk structure of $MoS_2$ consists of 2D, 3-atom thick nanosheets stacked using van der Waals forces that have a molybdenum (Mo) atom coordinated by six sulfur (S) atoms.[28] Specifically, $MoS_2$ nanosheets show intrinsically active sites (edge-unsaturated Mo sites and S vacancies), which exhibit favorable hydrogen adsorption energetics, while their basal planes remain inactive.[25,28–30] Despite the interest in the material, monolayer and few-layer $MoS_2$ still suffer from a limited density of catalytic sites, poor electronic conductivity, and mechanical fragility that complicates handling.[25,30,31] To address these limitations, recent efforts have focused on engineering edge terminations and controlling layer stacking in $MoS_2$ nanosheets assemblies.[25–27,30,32] In particular, the $MoS_2$ bilayer emerged as a promising high-performance electrocatalyst that serves as a robust and versatile platform.[25,30] Indeed, the bilayer can stabilize energetically favorable closed-edge reconstructions and enable targeted defect modulation on the basal plane, leading to band-gap reduction, electronic state rearrangement, and optimized active sites that improve hydrogen adsorption thermodynamics.[25,30]

Here, we synthesized and characterized $MoS_2$ bilayers hybridized with reduced graphene oxide (rGO), which was strategically incorporated with different levels of reduction to mainly tune the system's electronic conductivity, adjust the $MoS_2$ interlayer spacing, and potentially enhance the electrocatalytic activity. To this end, $MoS_2$ bulk was exfoliated in a liquid medium, and the resulting material was combined with an rGO dispersion before drop-casting to fabricate the electrodes. Linear-sweep voltammetry was employed to evaluate the performance of the resulting electrodes in the hydrogen evolution reaction (HER). The MoS$_2$ and rGO materials were physically characterized by scanning-tunneling microscopy (STM), whereas their chemical properties were evaluated by ultraviolet-visible (UV-Vis) spectroscopy, confirming the successful synthesis of MoS$_2$ bilayers and rGO nanosheets. The results demonstrated that MoS$_2$ bilayers hybridized with rGO under a 2-hour moderate reduction electrocatalytically outperformed the pristine MoS$_2$ bilayer and other rGO reduction conditions, exhibiting the lowest overpotential (*ca.* 70 mV) and Tafel slope (46 mV dec$^{-1}$). Furthermore, the MoS$_2$ bilayers hybridized with rGO exhibit a promising overpotential for a range of higher current densities compared to state-of-the-art studies involving MoS$_2$ and rGO, as well as to a commercial 20% platinum on carbon (Pt/C) electrode.[24,26,33–36] These findings indicate that controlling the combination of MoS$_2$ bilayers with moderately reduced rGO nanosheets may prevent further restacking of nanostructures, improve conductivity, retain oxygenated groups to increase interlayer spacing, and maximize the exposure of electroactive edge sites – enhancing the efficiency of HER.

**RESULTS AND DISCUSSION**

MoS$_2$ bilayers were synthesized through liquid exfoliation, as illustrated in Figure 1a.[37] Initially, the MoS$_2$ powder was dispersed in isopropanol using magnetic stirring, as shown in Figure 1a.1. The exfoliation process was then performed in an ultrasonic bath, which generated cavitation bubbles that collapsed, producing high-speed liquid microjets and shock waves. This process yielded shear stresses and tensile forces in the van der Waals gaps of MoS$_2$, facilitating the separation of the bulk material into MoS$_2$ nanosheets (Figure 1a.2).[38] The exfoliated MoS$_2$ was subsequently isolated using centrifugation, which allowed larger particles to settle more effectively. Meanwhile, the graphene oxide was dispersed in deionized water and heated at 80°C under magnetic stirring. Ammonium hydroxide (NH$_4$OH) was added until the pH reached *ca.* 10 (Figure 1b). The dispersion was maintained under these mild alkaline thermal conditions for 1, 2, or 4 h, resulting in partially reduced rGO. Aliquots of the dispersion were collected at specific time intervals to prepare hybridized samples with varying reduction degrees of rGO (Figure 1.b1). The electrodes were fabricated using a drop-casting technique with a dispersion of 90 weight-percent (wt%) MoS$_2$ and 10 wt% rGO (Figure 1.b2).

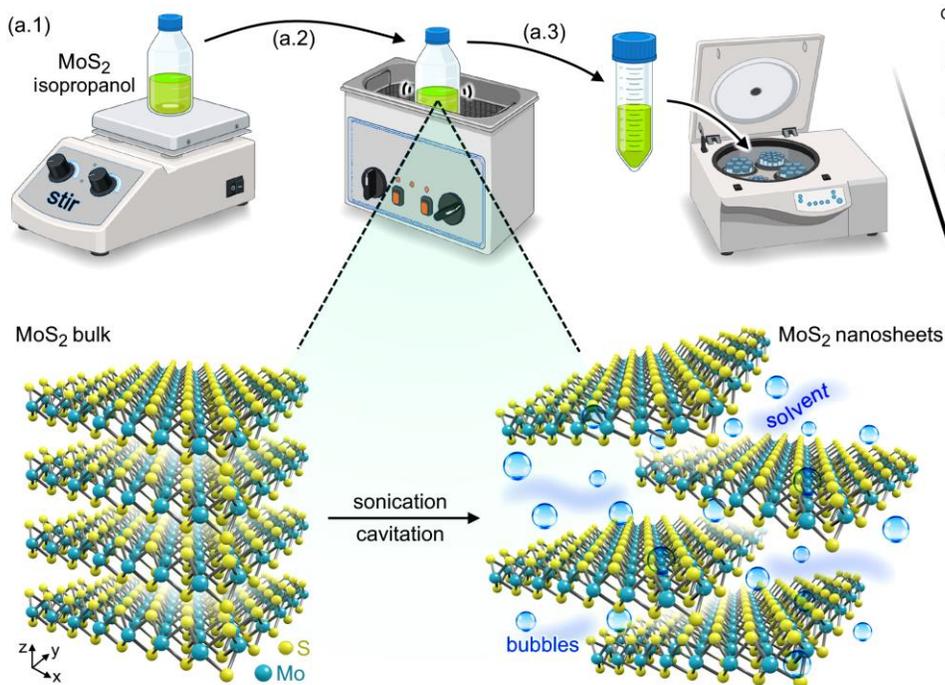
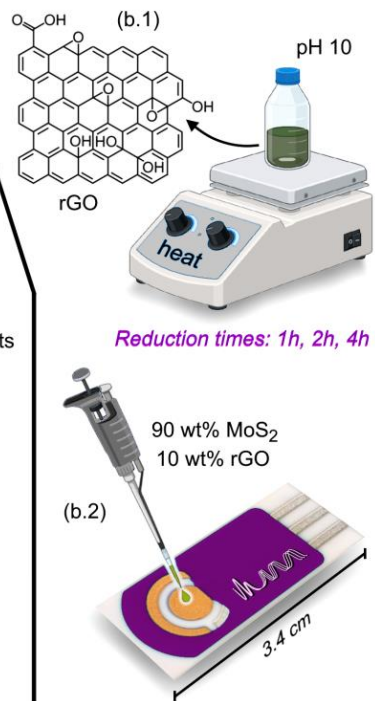
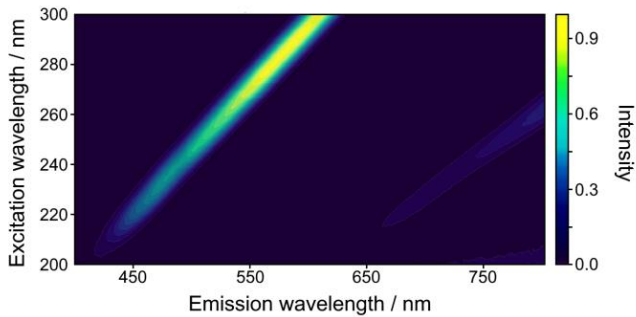
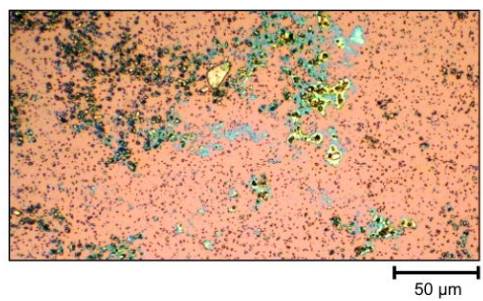

**Figure 1: MoS₂ nanosheets and rGO samples: preparation and initial characterization. (a)** Liquid exfoliation of (a.1) MoS₂ bulk dispersed in isopropanol. (a.2) Sonication and cavitation processes leading to MoS₂ nanosheets. (a.3) Centrifugation was used to settle the larger (non-exfoliated) particles of MoS₂. **(b)** Thermal reduction of graphene oxide under alkaline conditions. (b.1) Chemical structure of rGO, and reduction times investigated in this study: 1 h, 2 h, and 4 h. (b.2) Drop-casting of MoS₂@rGO (90:10 wt%) suspension onto the working electrode. Illustrations in (a,b) were created using BioRender. **(c)** Fluorescence spectrum and **(d)** optical microscopy image of liquid-exfoliated pristine MoS₂.

The fluorescence spectrum in Figure 1c showed a predominant emission in the 550-620 nm range, which revealed the presence of MoS₂ nanosheets after liquid exfoliation. Unlike bulk material, MoS₂ nanosheets emit light strongly due to two-dimensional quantum confinement occurring across material thickness. This happens because, as the thickness decreases, the indirect bandgap, which has lower energy than the direct bandgap in the bulk material, shifts upward, eventually leading to a transition to a direct-gap material in a single nanosheet.[30,39] However, the optical microscopy image in Figure 1d revealed a high degree of polydispersity in the pristine $MoS_2$ after undergoing liquid exfoliation. The bluish areas indicate the presence of $MoS_2$ nanosheets, while the yellowish regions represent the bulk material.[40]

In Figure 2a, the STM image of the $MoS_2$ nanosheets revealed that they were *ca.* 10-30 nm wide, consistent with the range reported in the literature.[38] The power level of the ultrasonic bath can induce bubble collapse near the crystals, leading to edge fragmentation and a reduction in sheet size. Furthermore, the height profile obtained from the STM image indicated that the as-fabricated $MoS_2$ nanosheets exhibit a thickness of *ca.* 1.1 nm (Figure 2b). This observation can be clarified by understanding that a freestanding $MoS_2$ monolayer displays a thickness of approximately 0.3 nm. However, when it's placed on a substrate, the apparent thickness increases to about 0.6 nm, which arises from the van der Waals separation between the monolayer and the substrate.[24] Thereby, the thickness of *ca.* 1.1 nm (Figure 2b) suggests that our synthesis protocol produced $MoS_2$ nanosheets comprising consistent bilayer structures.

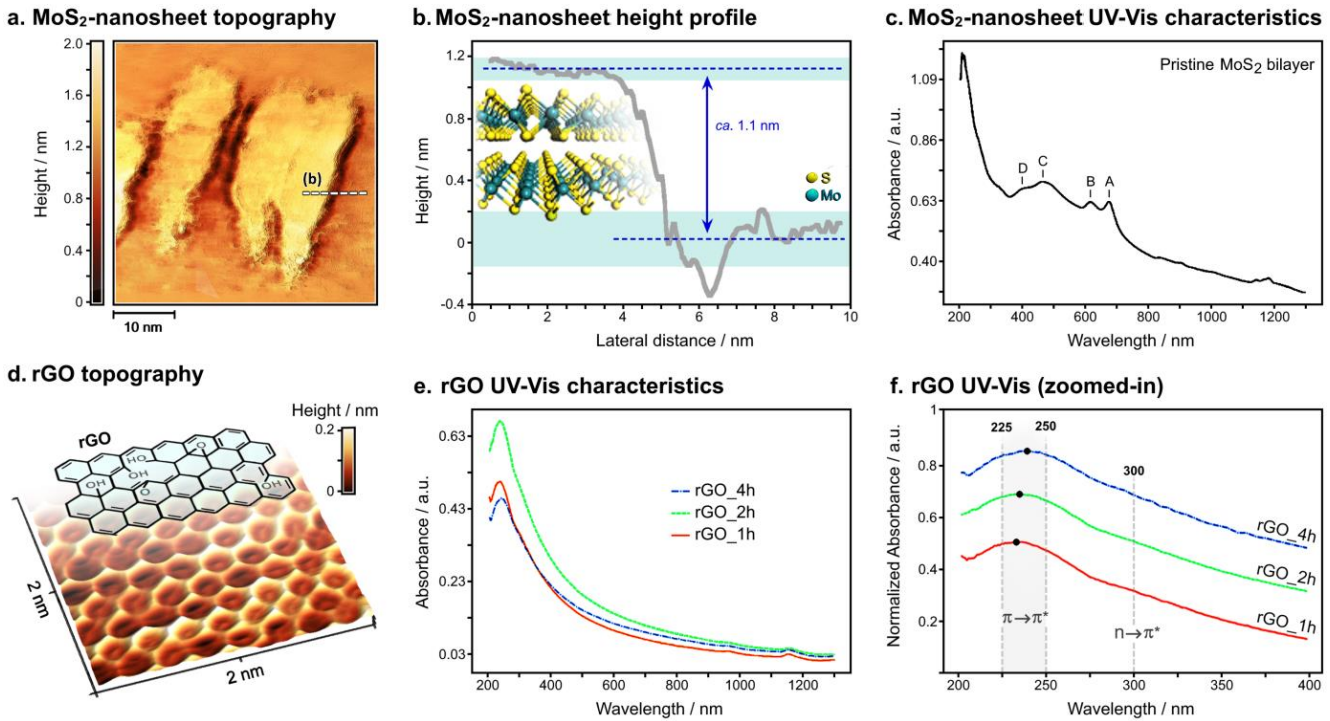

**Figure 2: Characterization of pristine $MoS_2$ and rGO with different degrees of reduction. (a)** STM topography acquired for the $MoS_2$ nanosheets. **(b)** Height profile acquired along the dashed line depicted in panel (a). The inset illustrates the $MoS_2$ bilayer structure. **(c)** UV-Vis spectrum of pristine $MoS_2$ bilayer. **(d)** Three-dimensional plot of the STM topography acquired for the rGO. A partially reduced rGO layer is illustrated on the top, corresponding to the topography map. **(e)** UV-Vis spectra of rGO at different degrees of reduction. **(f)**

Highlight (zoom-in) of the rGO absorption bands in the 200-400 nm wavelength range. Each curve was normalized to its own maximum absorbance.

The UV-Vis spectrum of exfoliated pristine $MoS_2$ bilayers shows four characteristic peaks of transition metal dichalcogenides with trigonal prismatic geometry (Figure 2c).[41,42] The absorption peaks A (*ca.* 674 nm) and B (*ca.* 613 nm) correspond to direct excitonic transitions at the K points of the Brillouin zone.[39,41–43] These transitions arise from direct electronic excitations between the spin-orbit-split valence bands and the conduction band, which only occur in few-layered $MoS_2$ due to its direct bandgap nature.[39,41–43] This phenomenon contributes to its strong light-matter interaction, as illustrated in the fluorescence spectrum of Figure 1c.[39] In contrast, the C (*ca.* 462 nm) and D (*ca.* 398 nm) absorption peaks are associated with fully quantum-confined $MoS_2$ domains, which exhibit higher optical bandgap energies.[44] As expected for quantum confinement, the energies of these peaks show size-dependent behavior in $MoS_2$ nanocrystals, with a progressive blueshift as the particle size decreases.[43,45] The energy of each peak ($E_{(peak)}$) was calculated using $E_{(peak)} = h \times c \times \lambda^{-1}$, in which $h$ is Planck's constant ($6.626 \times 10^{-34}$ J s), $c$ is the speed of light ($2.998 \times 10^8$ m s$^{-1}$), and $\lambda$ is the peak wavelength. The calculated energies were $E_{(A)} = 1.84$ eV, $E_{(B)} = 2.02$ eV, $E_{(C)} = 2.68$ eV, and $E_{(D)} = 3.12$ eV. The energy difference for excitonic transitions, $E_{(B)} - E_{(A)} = 0.18$ eV, suggests the presence of $MoS_2$ bilayers, aligning with the STM image of Figure 2a and first-principles predictions from literature.[46]

Direct evidence of structural disorder in the rGO nanosheet was obtained through atomic-scale STM imaging (Figure 2d). The topography width reconstruction showed a peak-to-peak distance of 0.24 - 0.25 nm, similar to the periodicity observed in defect-free highly oriented pyrolytic graphite (HOPG) substrates.[47] From height mapping, variations of up to 0.2 nm were observed, attributed to the atomically thin layer thickness of rGO. Additionally, the rGO nanosheet is characterized by a lack of long-range atomic-scale order. The dominant features appear as atomic-sized spots arranged into slightly deformed domains. These domains form clearly ordered patterns of a few nanometers in size (as seen in the midsection of Figure 2d), while in other areas, local order is less apparent (such as the bottom right corner of Figure 2d). This suggests that regions with some structural imperfections coexist with relatively ordered areas. The disorder may stem from oxygen functionalities covalently attached to the basal plane of the graphene sheet, which can persist in substantial quantities even after chemical reduction, as shown in the chemical structure (Figure 2d-top).[47] These results align with those of Paredes *et al.*, who used STM along with complementary Raman and X-ray photoelectron spectroscopy analyses to study chemically reduced graphene oxide.[47] Moreover, Szabó *et al.* proposed a structural model for graphite oxide in which the carbon skeleton of graphene layers becomes highly corrugated due to the attachment of large quantities of oxygen functional groups.[48] This corrugation or deformation of the graphene lattice likely persists to some extent even after removing most of the oxygen via chemical reduction. Obviously, overcoming an energy barrier and likely some thermal annealing would probably be necessary to restore a perfectly flat, pristine graphene structure from the deformed one. Thereby, in our experiments, we assert that both atomic vacancies and lattice corrugation are probably present in our rGO nanosheets, likely causing the local distortions observed in Figure 2d.

The rGO dispersions at various reduction degrees were analyzed using UV-Vis spectroscopy (Figure 2e), revealing characteristic bands in the 200-300 nm range, consistent with the literature.[47,49,50] The rGO_2h sample displayed the highest absorbance, followed by rGO_1h and rGO_4h. We attribute this observation to the eventual distinct concentrations of each evaluation dispersion prepared for the UV-Vis characterization. Figure 2f exhibits each rGO spectrum normalized to its own maximum absorbance. The UV-Vis spectrum of graphene oxide typically exhibits peak absorption within the 225-250 nm range and a shoulder at *ca.* 300 nm.[47,50] In this work, the redshift observed in the grey region of Figure 2f, from *ca.* 233 nm (rGO_1h) to *ca.* 238 nm (rGO_4h), is related to the $\pi \rightarrow \pi^*$ transition and suggests an increase in electronic conjugation ascribed to the C=C bond.[42,51] The shift occurs due to the removal of oxygen-containing groups, which partially restores the $sp^2$-conjugated carbon network. The slight shoulders may also be observed in Figure 2f (centered at 300 nm). We attributed their occurrence to the $n \rightarrow \pi^*$ transitions of C=O bonds.[47,50] As the material is reduced, the removal of oxygen leads to a decrease in the absorbance shoulder, as visualized in Figure 2f for 1, 2, and 4-hour reduction protocols. Consequently, the progression of redshift at the 225-250 nm wavelength, along with the decreasing shoulder at 300 nm, corroborates that our synthesis protocol resulted in distinct degrees of reduction of rGO nanosheets.

The HER efficiency of the $MoS_2$ bilayer hybridized with rGO was assessed by linear sweep voltammetry in 1 M sulfuric acid ($H_2SO_4$) solution, as shown in Figure 3a. The working electrodes were modified with the investigated bilayer nanosheets, *i.e.*, pristine $MoS_2$, $MoS_2$@rGO_1h, $MoS_2$@rGO_2h, and $MoS_2$@rGO_4h, as illustrated in Figure 3a-inset. The voltammograms clearly indicate that the potential needed for HER was lower for $MoS_2$@rGO_2h compared to the other samples. Figure 3b shows the minimum overpotential ($\eta = \eta_0$) required to attain the reduction current density of $j_0 = 10$ mA cm$^{-2}$,[26] as well as the electrochemically active surface area (ECSA) estimated considering $j_0$. The elevated $\eta_0$ at the pristine-$MoS_2$ and $MoS_2$@rGO_4h samples suggests a deficiency of HER active sites and therefore a consequent low catalytic activity.[33] The $MoS_2$@rGO_1h and $MoS_2$@rGO_2h

samples exhibited the most favorable $\eta_0$ for achieving the HER condition at the studied electrode surfaces (Figure 3b). However, the lower ECSA of MoS$_2$@rGO_1h indicates that MoS$_2$@rGO_2h represents the most efficient nanosheet configuration for hydrogen generation through electrocatalysis in our investigation. This finding demonstrates that MoS$_2$@rGO_2h can achieve a significant improvement of $\eta_0$ while nearly preserving the ECSA of pristine MoS$_2$ bilayer thereby promoting a superior electron-transfer rate per unit area for HER.

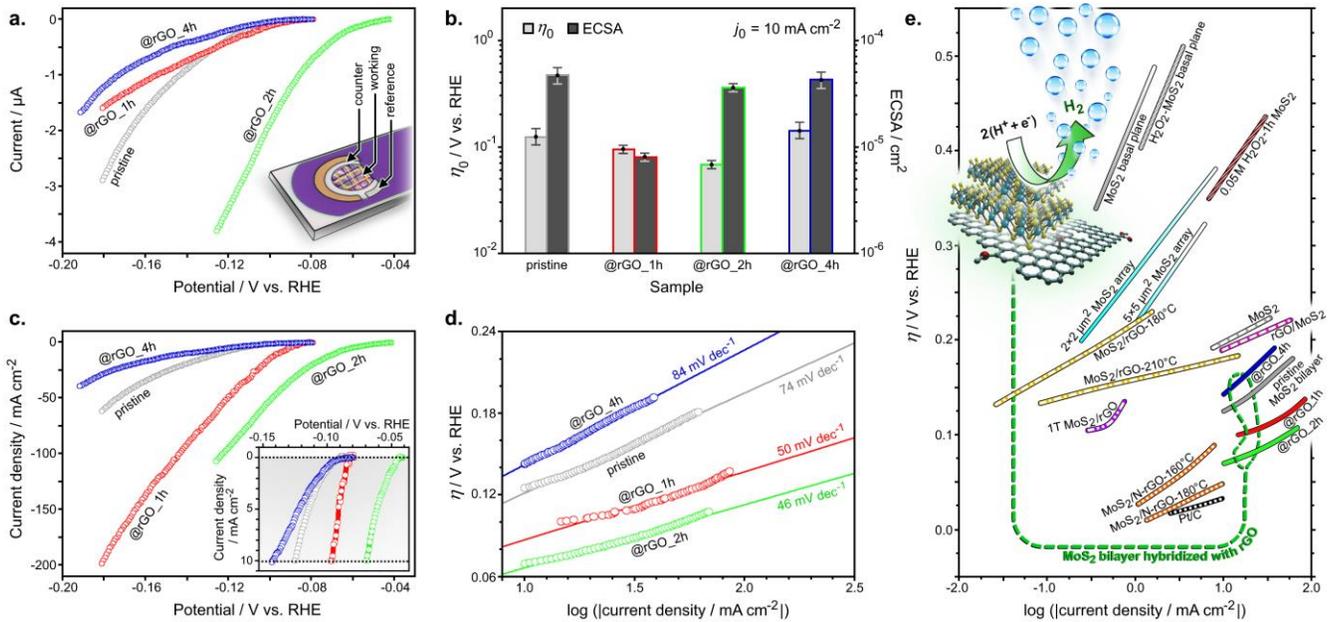

**Figure 3: HER measurements and characterization of MoS$_2$- and MoS$_2$@rGO-based bilayers.** **(a)** Linear sweep voltammograms recorded in 1 M H$_2$SO$_4$ at 2 mV s$^{-1}$. The inset illustrates the modified working electrode for the measurement. **(b)** $\eta_0$ and ECSA for each sample. **(c)** Polarization curves, with an inset highlighting the current density behavior up to 10 mA cm$^{-2}$. **(d)** Tafel plots, including their respective slopes estimated from linear fitting, with R$^2 \geq 0.98$. **(e)** HER performance comparison: MoS$_2$ bilayer hybridized with rGO (this work), MoS$_2$,[33,35] rGO/MoS$_2$,[35] MoS$_2$ arrays,[33] hydrogen peroxide (H$_2$O$_2$)-treated MoS$_2$,[26] temperature-treated MoS$_2$/rGO,[36] 1T phase MoS$_2$,[34] MoS$_2$/nitrogen (N)-doped rGO,[24] and a commercial 20% Pt/C electrode.[24] The inset schematically illustrates the HER process in a structure composed of a MoS$_2$ bilayer and an rGO nanosheet.

The ECSAs calculated for the different experimental conditions were used to estimate the current density of the polarization curves, as depicted in Figure 3c. The behavior of the current density up to $j_0$ = 10 mA cm$^{-2}$ is plotted in Figure 3c-inset. Figure 3d shows the corresponding Tafel plots, along with their slopes estimated from the corresponding linear fits. The fit criteria were a coefficient of determination (R$^2$) higher than 0.98. The calculated Tafel slopes are (74 ±2) mV dec$^{-1}$, 50 ± 1 mV dec$^{-1}$, 46 ± 1 mV dec$^{-1}$, and 84 ± 2 mV dec$^{-1}$ for the pristine MoS$_2$ bilayer, MoS$_2$@rGO_1h, MoS$_2$@rGO_2h,

and MoS$_2$@rGO_4h, respectively. Pristine MoS$_2$ bilayers show an improved $\eta$ (*ca.* 120 mV) compared to the literature.[26,33,35] Recently, Shang *et al.*[30] demonstrated through first-principles calculations that under certain chemical environments, closed reconstructed edges represent the most thermodynamically stable configuration for MoS$_2$ bilayers. This reconstruction stabilizes dangling bonds of the edges, decreases the band gap, increases specific surface area, and enhances the density of states at the Fermi level. Our result is the experimental evidence that pristine MoS$_2$ bilayers may inherently provide a favorable platform for catalysis and energy-related applications, exhibiting enhanced HER activity, as predicted by the hydrogen adsorption Gibbs free energy of 73 meV estimated recently.[30] Regarding the characterization of MoS$_2$@rGO, the electrocatalytic performance proved to be dependent on the degree of reduction of graphene oxide. The decrease in the Tafel slopes upon material functionalization indicates superior kinetics in the HER process.[26] In addition, the MoS$_2$@rGO_2h hybrids exhibited the lowest $\eta$, outperforming pristine MoS$_2$ bilayers and hybrids composed of rGO reduced for 1 h and 4 h. Moderate graphene oxide reduction, as occurs with 1- and 2-h protocols, partially restores the sp$^2$ carbon network, improving electrical conductivity. It also retains some oxygenated groups, maintaining interlayer spacing and preventing nanosheet restacking. Conversely, extensive reduction, such as that lasting 4 h, can eliminate most oxygenated groups, increase hydrophobicity, and lead to strong π-π interactions, causing a loss of accessible surface area. Overall, these results highlight that the optimal degree of graphene oxide reduction was not the highest nor the lowest we assessed, but rather the intermediate level. This degree of reduction must be carefully controlled to enable a high synergy between the MoS$_2$ bilayers and rGO nanosheets, balancing conductivity, interlayer spacing, and nanosheet dispersion.[52]

The performance of the samples in the HER process was compared to the state-of-the-art studies on MoS$_2$ synthesized at various temperatures or sizes, MoS$_2$ modified with rGO, and a commercial 20% Pt/C electrodes (Figure 3e).[24,26,33–36] The inset schematically illustrates the HER process in a structure composed of MoS$_2$ bilayer and rGO nanosheet. The first step of the HER mechanism is the Volmer reaction, which involves the adsorption of hydrogen cations from the H$_2$SO$_4$ solution. The second is the desorption of molecular hydrogen from the cathode, via an electrochemical or a chemical pathway (Heyrovsky or Tafel step, respectively).[53–56] Kinetic models of HER suggest that the Tafel slope is related to the rate-determining step, with caveats such as local molecular hydrogen accumulation, and mass-transport limits to consider. Specifically, Tafel slopes of 120 mV dec$^{-1}$, 40 mV dec$^{-1}$, or 30 mV dec$^{-1}$ correspond to Volmer, Heyrovsky, or Tafel mechanisms, respectively.[53–56] Therefore, the HER for the MoS$_2$@rGO_2h sample follows a Volmer-Heyrovsky mechanism, which reinforces the enhancement of its electrocatalytic performance. In due course, Figure 3e shows that the MoS$_2$@rGO_2h hybrids exhibited the lowest $\eta$ value along with the broadest current density range (*viz.* 10-100 mA cm$^{-2}$), indicating a promising catalytic activity for HER when compared to current state-of-the-art materials. In addition, its $\eta$ value was 1.5 times greater than that of the commercial electrode, which has a current density range of 2-10 mA cm$^{-2}$.[24] Finally, it is also noteworthy that the pristine MoS$_2$ bilayer, produced using our protocol, exhibited better electrocatalytic activity for HER compared to other control samples from the literature (Figure 3e).[26,33,35] This reinforces the promising potential of MoS$_2$ bilayers as catalytic platforms for other hybrid 2D heterostructure assemblies envisioning more efficient energy conversion.

## CONCLUSION

MoS$_2$ bilayers were synthesized using liquid exfoliation and subsequently hybridized with rGO nanosheets at different degrees of reduction. The composition of the materials and the thickness of the MoS$_2$ nanosheets were characterized using fluorescence, STM and UV-Vis measurements, attesting of the presence of MoS$_2$ bilayers with a thickness of *ca.* 1.1 nm. Samples of pristine MoS$_2$ bilayers, MoS$_2$@rGO_1h, MoS$_2$@rGO_2h, and MoS$_2$@rGO_4h were tested as alternative low-cost electrocatalysts for HER. The pristine MoS$_2$ bilayers exhibited exceptional advantages for electrocatalytic hydrogen evolution, achieving an impressive overpotential of just *ca.* 120 mV and a Tafel slope of *ca.* 74 mV dec$^{-1}$ across the 10-100 mA cm$^{-2}$ range. This provides solid experimental evidence supporting the recent theoretical advances on MoS$_2$ bilayer systems. Such outcome, supported by the high-resolution STM topography of manifold MoS$_2$ bilayers, is poised to significantly enhance the understanding of MoS$_2$ bilayer edges and their critical role in influencing system's electrocatalytic activity. Concerning the MoS$_2$@rGO hybrid bilayers, our findings demonstrated that the electrocatalytic performance depends on the degree of reduction of the rGO nanosheets. Moderate reduction of graphene oxide helps prevent nanosheet restacking, enhances electrical conductivity by partially restoring the sp$^2$ carbon network, and retains oxygenated groups that optimize interlayer spacing and expose more electroactive edges. Notably, the MoS$_2$@rGO_2h hybrids exhibited superior performance compared to the pristine MoS$_2$ bilayer and the hybrids composed of rGO reduced for 1 h and 4 h, showing the lowest $\eta$ (*ca.* 70 mV) and reduced Tafel slope (46 mV dec$^{-1}$). In addition, the $\eta$ value for MoS$_2$@rGO_2h was only 1.5 times greater than that of a commercial 20% Pt/C electrode within a current density range of 10-100 mA cm$^{-2}$. This indicates that the optimal degree of rGO reduction for enhancing MoS$_2$ bilayer's performance in HER catalysis is not the highest, but rather an intermediate level. We argue that the moderate reduction facilitates synergy between these nanomaterials, optimizing conductivity, interlayer spacing, and the dispersion of MoS$_2$ bilayers, which are essential to enhance the HER activity of MoS$_2$ heterostructures. The application of pristine MoS$_2$ bilayer and rGO-hybridized MoS$_2$ bilayer to advance HER capabilities not only expands opportunities for strain engineering, defect manipulation, and high-end twist-angle assemblies based on 2D materials but also paves the way for future theoretical advancements aiming at providing atomistic elucidations of these materials systems, especially concerning their highly electroactive edges.

**MATERIALS AND METHODS**

**Synthesis of MoS$_2$.** A total of 300 mg of molybdenum(IV) sulfide powder (Sigma-Aldrich) was dispersed in 60 mL of isopropyl alcohol (Merck) while mixed with magnetic stirring. The resulting suspension was then sonicated for 90 min in a 100 W ultrasonic bath operating at 80 kHz and 50°C. Finally, the solution was centrifuged at 2000 rpm for 30 min to remove non-exfoliated particles.[57]

**Synthesis of rGO.** Graphene oxide powder was acquired from Sigma-Aldrich. A total of 3.225 mL of 2 g L$^{-1}$ graphene oxide solution was mixed with 16.775 mL of ultrapure water. The pH of the mixture was adjusted to 10 by adding 50 µL of NH$_4$OH (Sigma-Aldrich). The mixture was reduced at 80°C with continuous stirring, and 2 mL samples were collected at 1, 2, and 4 h to isolate rGO with different reduction levels.

**Preparation of MoS$_2$@rGO dispersions.** MoS$_2$@rGO dispersions with a mass ratio of 90:10 and different levels of rGO reduction (1, 2, and 4 h) were prepared. First, the synthesized MoS$_2$ and rGO dispersions were dried. Then, the resulting materials were dispersed in isopropanol in the desired proportion.

**Electrode preparation.** Screen-printed gold working and counter electrodes, along with a silver reference electrode (Metrohm) were used for electrochemical measurements. The working electrode was prepared by drop-casting 5 µL of the MoS$_2$@rGO suspension, with varying degrees of rGO reduction. A nitrogen jet was applied directly to the electrode to facilitate drying.

**Optical microscopy.** MoS$_2$ nanosheets dispersion (*ca.* 20 µL) was dropped on silicon dioxide wafers, vacuum dried, and then visualized in the optical microscope with ×100 magnification in reflection mode (Motic BA310Met).

**STM characterization.** STM images were acquired using a Nanosurf NaioSTM operating at constant voltage (50.1 mV), Setpoint = 1 nA, Adaptive PID, and Const. Drive mode, with the scanning probe tip cut from Pt/iridium wire (as provided by Nanosurf). Dry MoS$_2$ nanosheets were prepared by drop casting (4 µL) on HOPG substrates, using the as-prepared MoS$_2$ dispersion after centrifugation to remove non-exfoliated particles. The MoS$_2$ samples were dried out in a low vacuum chamber at $10^{-1}$ - $10^{-2}$ mbar for 48 hours before the STM analysis. Dry rGO samples were prepared by drop casting (4 µL) on a HOPG substrate, using the as-prepared rGO_4h dispersion collected after a *ca.* 1-hour stirring. The rGO samples were dried out in a low vacuum chamber at $10^{-1}$ - $10^{-2}$ mbar for 48 hours before the STM analysis.

**UV-Vis spectroscopy.** A Shimadzu UV-Vis spectrophotometer (UV-2600i) equipped with the optional ISR-2600Plus Integrating Sphere was used to obtain the absorption spectra of the synthesized MoS$_2$ and rGO dispersions. The samples were scanned using a quartz cuvette with a 1 cm optical path over a wavelength range of 200 nm to 1300 nm with a 0.5 nm step size. In the MoS2 and rGO analyses, respectively, isopropyl alcohol and NH$_4$OH aqueous solution were employed for calibration and background correction.

**HER characterization.** The electrocatalytic performance for hydrogen evolution was studied using linear sweep voltammetry at a scan rate of 2 mV s$^{-1}$, with a step of 1 mV, in a 1 M H$_2$SO$_4$ solution purged with nitrogen gas for 30 minutes before the measurements.

## ACKNOWLEDGMENTS

We acknowledge the Brazilian Council for Scientific and Technological Development (CNPq), processes 447294/2024-5, 312243/2025-1, 315795/2025-5, and 135092/2025-6. L.M. acknowledges the São Paulo Research Foundation (FAPESP) through the Research Center of Molecular Engineering for Advanced Materials (CEMol), process number 24/00989-7. L.M. is grateful to Renato S. Lima, Flávio L. de Souza, and Hudson G. Zanin for their active participation in the Advanced Laboratory lectures on Energy and Sustainability at Ilum/Brazilian Center for Research in Energy and Materials. M.S. thanks FAPESP (24/15173-2 and 23/17576-4). The authors are grateful to Dr. Valéria S. Marangoni, Alessandro S. Mourato, Rhuan D. M. de Assis, and Aline C. Dias for technical support. This article is dedicated to the memory of the late Prof. Dr. Rogério Cezar de Cerqueira Leite, who made significant contributions to Brazil's science and technology ecosystem, and the late Prof. Dr. Antonio Rubens Britto de Castro, whose legacy lives on through the many professionals inspired by his example of rigor, creativity, and dedication to Brazilian science.

## CONFLICT OF INTEREST

The authors declare no conflict of interest.

## AUTHOR CONTRIBUTIONS

B.F.B., G.F., J.P.S.M., and L.F.M.R. contributed equally to this work. B.F.B., G.F., J.P.S.M., and L.F.M.R. synthesized the $MoS_2$ and hybrid $MoS_2$@rGO nanomaterials, performed optical microscopy, STM, and UV-Vis characterization, and revised the manuscript. G.D.V. synthesized the rGO samples and revised the manuscript. C.D.F. provided the STM calibration to measure the HOPG substrates and revised the manuscript. M.S. discussed the results and revised the manuscript. L.M.M.F. prepared the figures, discussed the results, wrote the first draft, and revised the manuscript. L.M. supervised all the experiments, performed HER measurements, analyzed the experimental data, prepared the figures, and wrote/revised the manuscript.

## DATA AVAILABILITY STATEMENT

The data that support the findings of this study are available from the corresponding author upon reasonable request.

## AUTHOR INFORMATION
**- E-mail**

Leandro Merces: leandro.merces@ilum.cnpem.br

**- ORCID**

Bruno Ferreira Brischi: 0009-0005-8779-5276

Gabriela Frajtag: 0000-0002-9161-3114

João Pedro da Silva Mariano: 0009-0001-3456-4475

Laís Fernanda Medeiros Ruela: 0009-0007-0965-3939

Gustavo Duarte Verçosa: 0000-0003-0440-0645

Camila Desiderio Fernandes: 0009-0005-0517-6907

Murilo Santhiago: 0000-0002-9146-9677

Letícia Mariê Minatogau Ferro: 0000-0002-2701-618X

Leandro Merces: 0000-0002-6202-9824